\RequirePackage{amsmath} 
\documentclass{llncs}
\usepackage{algorithm,amsfonts,amsmath,amssymb,bm,verbatim}
\usepackage{cellspace,chngcntr,colortbl,epsfig,graphicx,mathtools,xr-hyper,hyperref,multirow,tabularx,tikz,xcolor}
\usepackage{arydshln}
\usepackage[misc]{ifsym}
\usepackage[all]{xy}
\usepackage[noend]{algpseudocode}
\usepackage[normalem]{ulem}

\spnewtheorem{assumption}[theorem]{Assumption}{\bfseries}{\itshape}
\spnewtheorem{prop}[theorem]{Proposition}{\bfseries}{\itshape}
\spnewtheorem{rem}[theorem]{Remark}{\bfseries}{\itshape}

\DeclarePairedDelimiter\bra{\langle}{\rvert}
\DeclarePairedDelimiter\ket{\lvert}{\rangle}
\DeclarePairedDelimiterX\braket[2]{\langle}{\rangle}{#1 \delimsize\vert #2}

\newcommand{\algrule}{\Statex\par\vskip2pt\hrule\par\vskip-2pt}
\algrenewcommand\algorithmicrequire{\textbf{Input:}}
\algrenewcommand\algorithmicensure{\textbf{Output:}}

\graphicspath{{../figures/}{./figures/}{../../figures/}}
\makeatletter
\def\input@path{{./texts/}}
\makeatother 

\begin{document}

\title{New directions in dynamical expectation estimation}

\author{Panjin Kim \and Kyung Chul Jeong \and Yun-Tak Oh$^{(\text{\Letter})}$}
\institute{The Affiliated Institute of ETRI, Daejeon 34044, Korea \newline ytak0105@nsr.re.kr}


\maketitle
\begin{abstract}
Computing dynamical expectation values typically relies on approximations optimized for the state or observable separately, without accounting for how their errors combine in the final expression.
This work explores an alternative approach in which each approximation is guided by both the state and the observable, accounting for how they jointly determine the target expectation value.
This idea is implemented through a sweep algorithm with coupled loss functions for forward state and backward observable updates.
Exact error relations provide an analytical rationale for how the proposed losses can improve the accuracy.
Numerical tests on 30-qubit random circuits show errors two to three orders of magnitude smaller than those of variational state compression at equal bond dimensions.
These results motivate further exploration of joint state and observable approximation for dynamical expectation value estimation.
\end{abstract}

\section{Introduction}\label{sec:intro}
Computing expectation values of observables is a foundational task throughout physical sciences, including quantum information.
Due to the hardness of classically simulating dynamics of quantum systems\;\cite{feynman82,jozsa10,cirac07}, practical calculations generally rely on controlled approximations.
Representative approaches include approximating quantum states with tensor networks such as matrix product states (MPS), as well as approximating observables in the Heisenberg picture using Pauli propagation or Heisenberg-picture DMRG\;\cite{dmrg,vidal03,mps,gottesman98,pauliprop,hdmrg}.

Although such methods have firm theoretical foundations for approximating evolved \textit{states} or \textit{observables}, their objectives are not necessarily well suited to computing the \textit{expectation values}.
Their approximation criteria treat the state or observable \textit{in isolation}, without accounting for the object with which it is ultimately computing.

The central message of this work is that approximations of states and observables should be guided by how their errors combine in the final expectation value.
Even though the optimal implementation of this idea remains open, one choice explored in this work is a loss of the following form:
\begin{align*}
    \ell\!\left(\rho;\rho^\star,\bar{O}\right)
    &=
    \gamma
    \left|
        \operatorname{Tr}\!\left[
            \left(\rho-\rho^\star\right)\bar{O}
        \right]
    \right|^2
    +
    \mu
    \left\|\rho-\rho^\star\right\|^2.
\end{align*}
Here, $\rho$ is the candidate approximation, $\rho^\star$ is the uncompressed target, and $\bar{O}$ is the observable propagated backward and held fixed during the update.
The second term measures the conventional state-only approximation error, whereas the first measures the resulting error in the expectation value of $\bar{O}$.
Their relative influence is controlled by $\gamma$ and $\mu$.
This expression is presented as an example rather than a uniquely preferred objective.
Its motivation and the complete loss construction are developed in Section\;\ref{sec:algorithm} and Section\;\ref{sec:rationale}.

Although the approach developed here---a sweep-like optimization along the time direction---is not claimed to be the optimal realization of this principle, the error relations derived provide a concrete analytical rationale: minimizing state and observable errors separately may preserve features of little relevance to the expectation value, whereas aligning the two can prioritize the features that matter most.
Consistent with this rationale, numerical experiments using the loss function considered here yield errors several orders of magnitude smaller than those produced by MPS-based circuit simulations and Pauli propagations in the tested settings. 
\section{Existing Approximation Methods}\label{sec:background}
In MPS-based simulations, evolution is interleaved with truncations that restrict the bond dimension, typically through truncated singular value decomposition or variational compression\;\cite{vidal04,TEBD,schollwock11}.
At step $t$, the compressed state $\ket{\tilde{\psi}_t}$ for the uncompressed target $U_t\ket{\psi_{t-1}}$ is chosen to minimize
\begin{align*}
    \epsilon_t
    =
    \left\|
        U_t\ket{\psi_{t-1}}
        -
        \ket{\tilde{\psi}_t}
    \right\|_2,
\end{align*}
subject to the controlled bond dimension limit.
For normalized pure states, this objective can equivalently be expressed as maximizing fidelity\;\cite{stoudenmire23}.

In a broad sense, Pauli propagation is the Heisenberg picture counterpart of state evolution.
Rather than evolving the state forward, it propagates the final observable backward through the evolution.
An observable expanded in Pauli strings can be represented as a tensor network, with each local physical index labeling $I$, $X$, $Y$, or $Z$ and the virtual bonds encoding the coefficients of the resulting Pauli strings.
The expectation value is obtained by contracting the observable tensors with the MPS and its Hermitian conjugate through the local Pauli basis tensor
\begin{align*}
    P_{ss'\mu}
    =
    (\sigma_\mu)_{ss'},
    \qquad
    \sigma_\mu\in\{I,X,Y,Z\},
\end{align*}
where $\mu$ is the Pauli index and $s$, $s'$ connect to the MPS and its Hermitian conjugate, respectively.
Two common strategies for truncating an evolving observable are Schmidt truncation of the matrix product operator (MPO)  minimizing the Hilbert--Schmidt error, and discarding Pauli strings above a certain Pauli weight\;\cite{pauliprop,hdmrg}.
A recent hybrid approach combines forward MPS evolution with backward low-weight Pauli propagation and contracts the resulting approximations at an intermediate time\;\cite{mitm}.

Although these methods differ in detail, their truncation criteria remain single-sided: each truncation is determined from either the state or the observable alone.
Even in the meet-in-the-middle approach, the state and observable are truncated independently, without considering how their approximation errors combine in the final expectation value.

\section{Algorithm}\label{sec:algorithm}
An algorithm developed in this work is presented first, with its theoretical rationale deferred to Section\;\ref{sec:rationale}.
For simplicity, this paper focuses on expectation values arising from quantum dynamics, represented here as quantum circuits.
The same formulation applies whenever the dynamics can be decomposed into a sequence of local evolution maps.

Let $\mathcal{C}=U_L\cdots U_1$ denote a quantum circuit of $L$ unitary gates, where $U_t$ is the $t$-th gate.
Given the initial state $\rho_0=\ket{\psi_0}\!\bra{\psi_0}$ and an observable $O$, the goal is to approximate
\begin{align*}
    \operatorname{Tr}\!\left[
        \rho_0\mathcal{C}^\dagger O\mathcal{C}
    \right].
\end{align*}
Define the forward evolved state and backward evolved observable at slice $t$ as
\begin{align*}
    \rho_t
&=
    U_t\cdots U_1 \rho_0 U_1^\dagger\cdots U_t^\dagger,
    \\
    O_t
&=
    U_{t+1}^\dagger\cdots U_L^\dagger O U_L\cdots U_{t+1}.
\end{align*}
The initial ansätze $\tilde{\rho}_t^{(0)}$ and $\tilde{O}_t^{(0)}$ are assumed to be given at every time slice, with $\tilde{\rho}_0^{(0)}=\rho_0$ and $\tilde{O}_L^{(0)}=O$.\footnote{For example, these ansätze may be constructed using simple sweeps based on singular value decomposition.}
The superscript denotes the sweep index.
Let $\chi_\rho$ and $\chi_O$ denote the maximum virtual bond dimensions of the forward state and backward observable, respectively.

\begin{algorithm}[H]
	\small
	\caption{ }
	\begin{algorithmic}[1]
        \Require{$\mathcal{C} = (U_t)_{t=1}^{L}$, $\{\tilde{\rho}_t^{(0)}\}_{t=0}^L$, $\{\tilde{O}_t^{(0)}\}_{t=0}^L$, $\chi_\rho, \chi_O$}
		\Comment{$\tilde{\rho}_0^{(0)}=\rho_0$, $\tilde{O}_L^{(0)}=O$}
		\Ensure{$\langle \tilde{O} \rangle$}
		\algrule\vspace{2pt}
    \State $\tilde{\rho}_0^{(1)}\gets \tilde{\rho}_0^{(0)}$
    \For{$t=1,\ldots,L$}
    \Comment{Forward time sweep}
        \State $\displaystyle \tilde{\rho}_t^{(1)} \gets 
            \operatorname*{arg\,min}_{\rho:\,\chi(\rho)\leq\chi_\rho}
            \ell_F\!\left(\rho; U_t\tilde{\rho}_{t-1}^{(1)}U_t^\dagger, \tilde{O}_t^{(0)} \right)$
    \EndFor
    \State $\tilde{O}_L^{(1)}\gets \tilde{O}_L^{(0)}$
    \For{$t=L,\ldots,1$}
    \Comment{Backward time sweep}
        \State $\displaystyle \tilde{O}_{t-1}^{(1)} \gets
            \operatorname*{arg\,min}_{O':\,\chi(O')\leq\chi_O}
            \ell_B\!\left(O'; U_t^\dagger\tilde{O}_t^{(1)}U_t, \tilde{\rho}_{t-1}^{(1)} \right)$
    \EndFor
    \State $\langle\tilde{O}\rangle^{(1)} \gets
        \operatorname{Tr}\! \left[ \rho_0\tilde{O}_0^{(1)} \right]$
    \State \Return $\langle\tilde{O}\rangle^{(1)}$
	\end{algorithmic}
	\label{alg:main-algorithm}
\end{algorithm}
%
The two loss function $\ell_F$ and $\ell_B$ are defined by 
\begin{align}
    \ell_F\!\left(\rho;\rho^\star,\bar{O}\right)
    &=
    \gamma
    \left|
        \operatorname{Tr}\!\left[
            \left(\rho-\rho^\star\right)\bar{O}
        \right]
    \right|^2
    +
    \mu
    \left\|\rho-\rho^\star\right\|^2,
    \nonumber\\
    \ell_B\!\left(O;O^\star,\bar{\rho}\right)
    &=
    \gamma
    \left|
        \operatorname{Tr}\!\left[
            \bar{\rho}\left(O-O^\star\right)
        \right]
    \right|^2
    +
    \mu
    \left\|O-O^\star\right\|^2,
    \label{eq:loss-functions}
\end{align}
where the $\| \cdot \|$ is Hilbert--Schmidt norm and $\gamma$, $\mu$ are external parameters.
To clarify the role of each term, setting $\gamma=0$ reduces $\ell_F$ to conventional state compression, which minimizes the distance from the uncompressed target $\rho^\star$.
Turning on the $\gamma$-term additionally drives the approximate contraction $\operatorname{Tr}[\rho\bar{O}]$ toward its uncompressed counterpart $\operatorname{Tr}[\rho^\star\bar{O}]$.
The loss $\ell_B$ has the analogous interpretation with the roles of state and observable reversed.
Note that the updates obtained by minimizing the loss generally do not preserve unitarity, even though the underlying gates $U_t$ are unitary.

In the next section, it is shown how these loss functions can reduce expectation value errors relative to conventional methods.
\section{Error Relations and Rationale}\label{sec:rationale}
Assume that the initial approximations $\tilde{\rho}_t^{(0)}$ and $\tilde{O}_t^{(0)}$ are obtained by conventional SVD-based approximation\;\cite{TEBD}, corresponding to minimizing the loss functions with $\gamma=0$.
They yield two endpoint estimates, $\operatorname{Tr}\!\left[\rho_0\tilde{O}_0^{(0)}\right]$ and $\operatorname{Tr}\!\left[\tilde{\rho}_L^{(0)}O\right]$, either of which can serve as the starting point for the analysis.
For definiteness, let $b^{(0)}=\operatorname{Tr}\!\left[\rho_0\tilde{O}_0^{(0)}\right]$, and let $b^{(1)}$ denote the value returned by Algorithm~\ref{alg:main-algorithm}.
Writing the exact expectation value as $E$, the errors satisfy
\begin{align}\label{eq:error}
    E-b^{(i)} = R_1^{(i)} + R_2^{(i)},
    \qquad i=0,1,
\end{align}
where
\begin{align*}
    R_1^{(i)}
&=
    \sum_{t=0}^{L-1}
    \operatorname{Tr}\!\left[
        \tilde{\rho}_t^{(i)}d_t^{(i)}
    \right],
    \\
    R_2^{(i)}
&=
    \sum_{t=0}^{L-1}
    \operatorname{Tr}\!\left[
        \left(\rho_t-\tilde{\rho}_t^{(i)}\right)d_t^{(i)}
    \right],
    \\
    d_t^{(i)}
&=
    U_{t+1}^\dagger\tilde{O}_{t+1}^{(i)}U_{t+1}
    -
    \tilde{O}_t^{(i)}.
\end{align*}
Here, $d_t^{(i)}$ denotes the local error introduced by backward evolution at time slice $t$.
The term $R_1^{(i)}$ collects its contributions to the expectation value evaluated with the approximate states, while $R_2^{(i)}$ captures the combined effect of state approximation errors and backward compression errors.
For $i=0$, Eq.\;\eqref{eq:error} simply rewrites the conventional approximation error $E-b^{(0)} = \operatorname{Tr}\!\left[\rho_0\left(O_0-\tilde{O}_0^{(0)}\right)\right]$.

The following analysis explains the rationale for reducing the overall error in three steps:
\begin{itemize}
    \item The $\gamma$-terms penalize local changes in the expectation value, thereby reducing the magnitude of $R_1^{(1)}$.
    \item The remainder $R_2^{(1)}$ combines state and observable approximation errors and can be smaller than $R_1^{(0)}$ under suitable error-scaling conditions.
    \item The $\mu$-terms constrain directions left undetermined by the $\gamma$ terms alone, thereby regularizing the local optimization.
\end{itemize}

The first step is clear from the loss function. 
The $\gamma$-term in $\ell_B$ penalizes $|\operatorname{Tr}[\tilde{\rho}_t^{(1)}d_t^{(1)}]|^2$ at each time slice $t$, directly suppressing the individual contributions to $R_1^{(1)}$.

For the second step, both remainders satisfy
\begin{align}
    |R_2^{(i)}|
    \leq
    \sum_{t=0}^{L-1}
    \left\|\rho_t-\tilde{\rho}_t^{(i)}\right\|
    \left\|d_t^{(i)}\right\|,
    \qquad i=0,1.
\end{align}
Each contribution to $R_2^{(i)}$ is bounded by a product of two approximation errors, giving $R_2^{(i)}=O(L\epsilon^2)$ when both errors are $O(\epsilon)$.
In contrast, $R_1^{(0)}$ contains only one approximation error per term, through $d_t^{(0)}$, and is $O(L\epsilon)$ for bounded state norms.
For fixed $L$, both $R_2^{(0)}$ and $R_2^{(1)}$ are smaller in magnitude than $R_1^{(0)}$ for sufficiently small $\epsilon$, provided that its first-order contribution does not vanish.

Finally, notice that the \(\gamma\)-term penalizes deviations in only one scalar quantity, leaving approximation errors that do not affect this value unpenalized.
The \(\mu\)-term guides the choice among approximations yielding the same scalar value by favoring those closer to the uncompressed target. For this purpose, \(\mu\) need not be large compared with \(\gamma\).
The \(\mu\)-term also helps limit the residual norms entering the bound on \(R_2^{(i)}\).


\begin{figure}[htbp]
    \centering
    \includegraphics[width=0.98\linewidth]{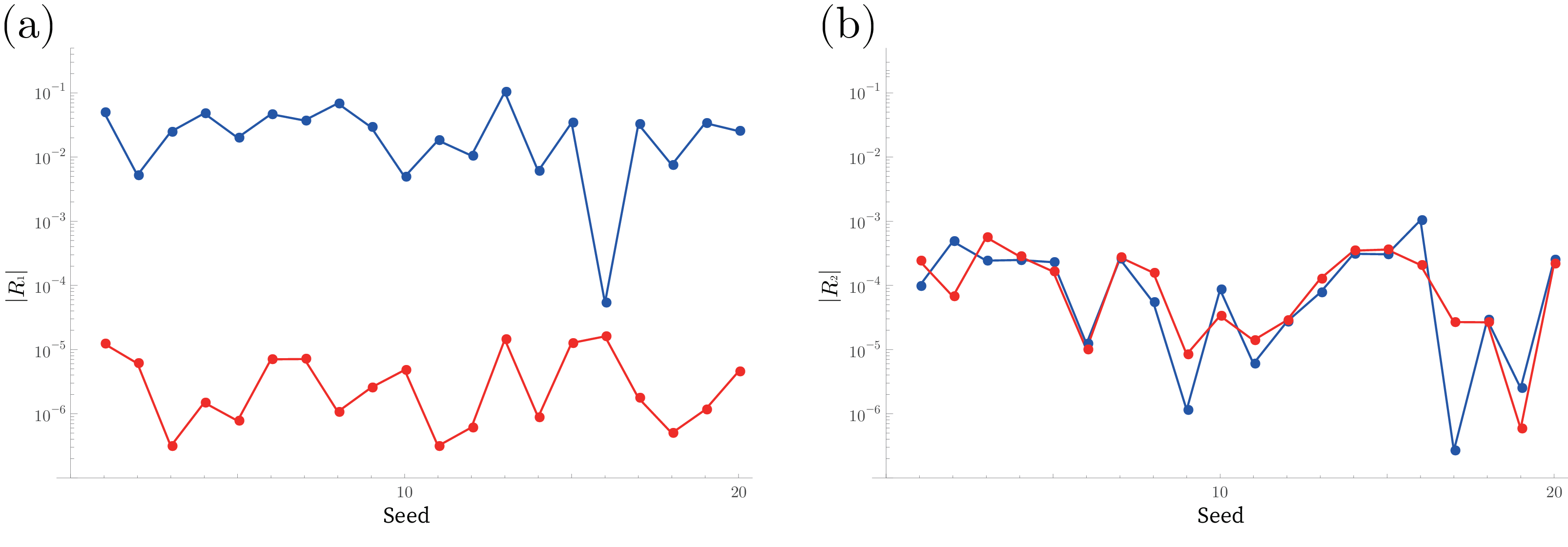}
    \caption{(a) Absolute values of $R_1^{(0)}$ (blue) and $R_1^{(1)}$ (red) for 20 barren plateau circuit instances, each with 20 qubits, depth 50 (see Section~\ref{sec:benchmark}), and maximum bond dimensions $\chi_\rho=\chi_O=30$. (b) Absolute values of $R_2^{(0)}$ (blue) and $R_2^{(1)}$ (red) for the same circuit instances and parameters as in (a).}
    \label{fig:errors}
\end{figure}

Figure~\ref{fig:errors} compares $|R_1^{(i)}|$ and $|R_2^{(i)}|$ for $i=0,1$ across 20 barren plateau circuit instances, each with $n=20$ qubits (see Section~\ref{sec:benchmark} for details on the circuits).
In these instances, Algorithm~\ref{alg:main-algorithm} suppresses $|R_1^{(1)}|$ relative to $|R_1^{(0)}|$ as intended, while both $|R_2^{(0)}|$ and $|R_2^{(1)}|$ generally remain smaller than $|R_1^{(0)}|$.

\section{Numerical tests}\label{sec:numerics}
This section numerically benchmarks the proposed time sweep algorithm against two representative conventional methods\;\cite{stoudenmire23,pauliprop}.

\subsection{Implementation details}
Section\;\ref{sec:algorithm} was intended to convey the central idea rather than implementation details, thus the specific realization is described here.
The state and observable in Algorithm\;\ref{alg:main-algorithm} are constructed by MPS.

\begin{figure}[htbp]
    \centering
    \includegraphics[width=0.98\linewidth]{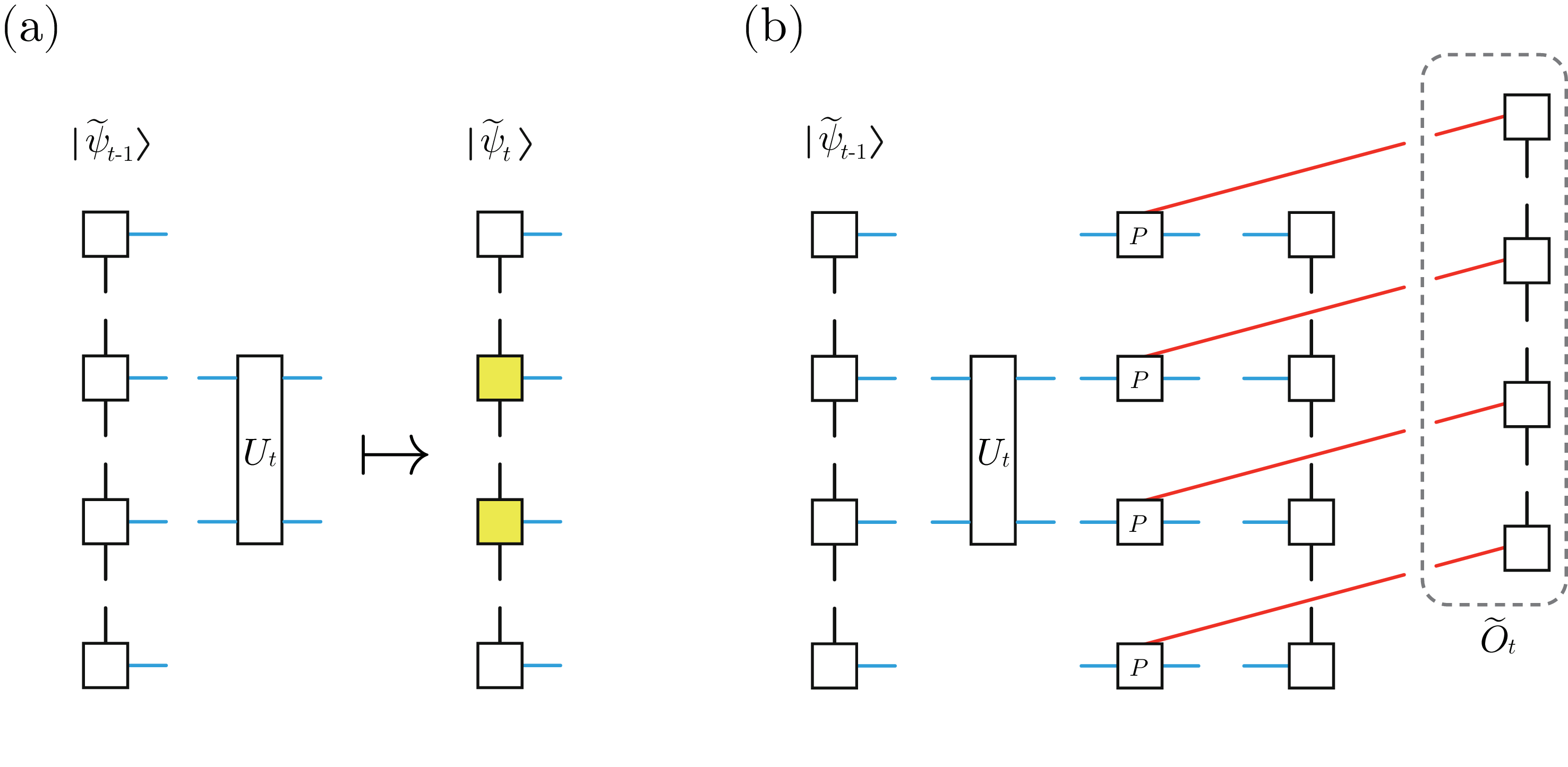}
    \caption{(a) Tensor network representation of one step of the forward time sweep with the blue bonds denoting the qubit indices. The yellow tensors represent the updated approximation to the evolved state up to canonicalization. The $\mapsto$ symbol between the networks denotes an update, which should not be interpreted as conventional state compression. State compression is considered only as the $\mu$-term in Eq.\;(\ref{eq:loss-functions}). (b) Tensor network representation of the target expectation value before the update, with $P$ denoting the local Pauli basis tensor and the red bonds denoting the Pauli indices. The optimization of the updated state $\ket{\tilde{\psi}_t}$ accounts for deviations from this target contraction through the $\gamma$-term in Eq.\;\eqref{eq:loss-functions}. Notice that the backward propagated observable $\tilde{O}_t$ is required to evaluate the forward loss $\ell_F$. Conversely, the forward evolved state is required to evaluate $\ell_B$ during the backward time sweep.}
    \label{fig:tn-forward-sweep}
\end{figure}

For brevity, only the forward time sweep is described below.
The backward sweep proceeds analogously, with the sweep direction reversed and the adjoint action of $U_t$ represented in the Pauli basis.
The core of the forward update is the minimization of $\ell_F$ in Eq.\;(\ref{eq:loss-functions}).
The $\mu$-term recovers conventional state compression, whereas the $\gamma$-term evaluates the error induced in the expectation value at time slice $t$ and therefore requires the backward propagated observable $\tilde{O}_t$.
Single-qubit gates can be absorbed exactly into an MPS without increasing its bond dimensions.
Multi-qubit gates, by contrast, can enlarge the virtual bond dimension and therefore require truncation when the prescribed limit is exceeded.
For simplicity, only nearest-neighbor two-qubit gates are considered here.
Holding $\tilde{O}_t$ fixed, the two MPS tensors on which $U_t$ acts are updated jointly, as illustrated in Fig.\;\ref{fig:tn-forward-sweep}(a).
Fixing either tensor while optimizing the other reduces the stationarity condition to a linear system for the tensor being updated.
After alternating between the two tensor updates for a few iterations, the sweep proceeds to the next gate.
For a fixed number of local update iterations (two in this work), the time and memory costs of a full sweep scale as $\operatorname{poly}(\chi_\rho,\chi_O,n,L)$, where $n$ is the number of qubits.

\subsection{Benchmarks}\label{sec:benchmark}

The benchmarks use the $n$-qubit barren plateau (BP) circuit introduced by McClean \textit{et al.}\;\cite{BP}.
An initial $R_y(\pi/4)$ rotation is applied to every qubit, followed by repeated layers of independently sampled single-qubit Pauli rotations and a nearest-neighbor controlled-$Z$ ladder.\footnote{Therefore, each layer contains $n$ single-qubit rotations and $n-1$ nearest-neighbor controlled-$Z$ gates.}
The circuit is illustrated in Fig.~2(b) of Ref.\;\cite{BP}.
At sufficient depth, the BP ensemble is expected to exhibit approximate unitary $2$-design behavior\;\cite{BP,BHH16}.

\begin{figure}[htbp]
    \centering
    \includegraphics[width=0.98\linewidth]{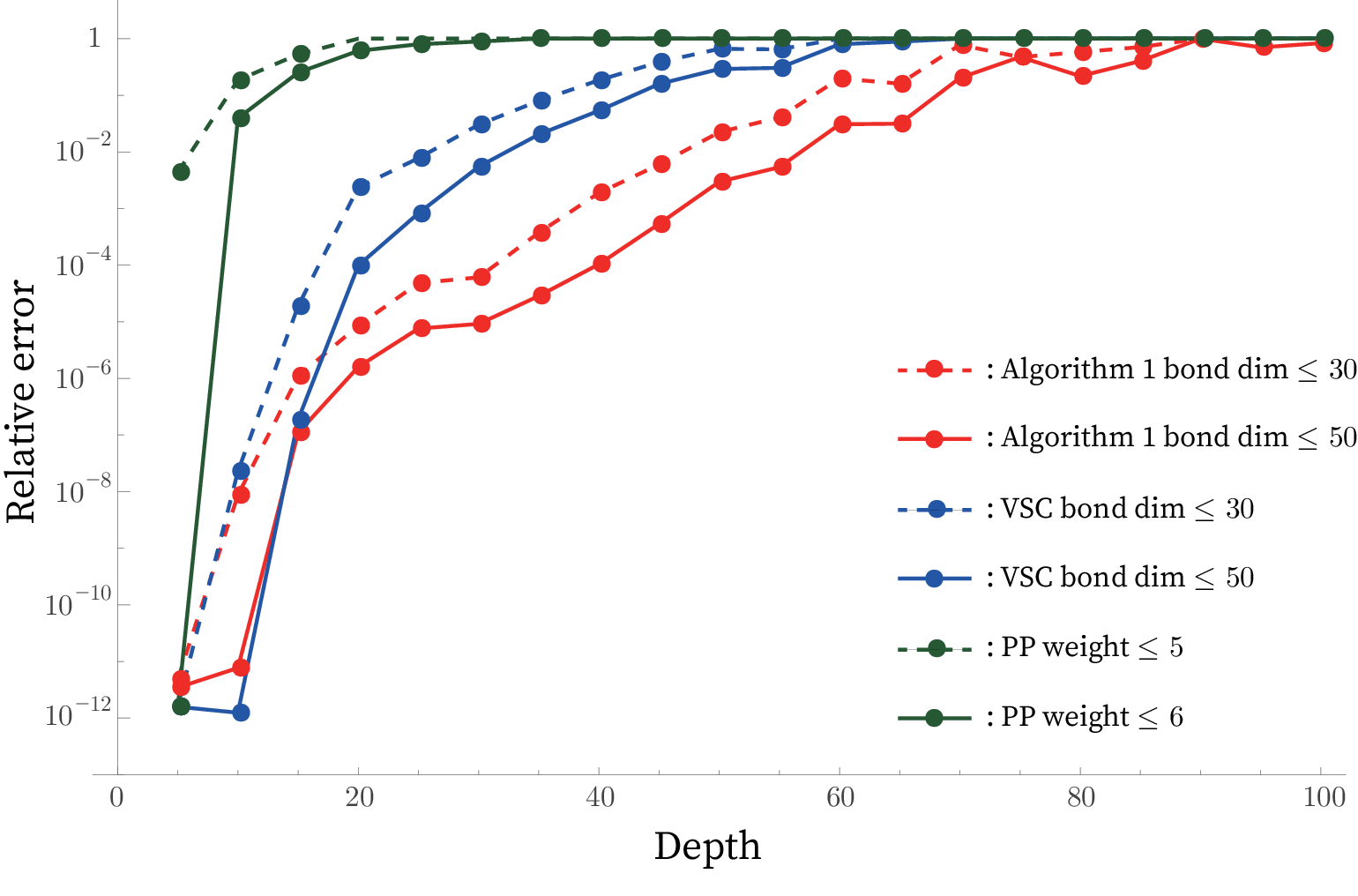}
    \caption{Relative errors of the three approximation methods, each averaged over the same 500 BP circuit instances with $n=30$ qubits. For Algorithm\;\ref{alg:main-algorithm}, the maximum bond dimension of the backward observable is set equal to that of the forward MPS. }
    \label{fig:benchmark}
\end{figure}

For $n=30$ qubits, 500 BP circuits are randomly generated with layer counts $L\in\{5,10,\ldots,100\}$.
Variational state compression (VSC)\;\cite{schollwock11,stoudenmire23}, Pauli propagation (PP)\;\cite{pauliprop}, and Algorithm\;\ref{alg:main-algorithm} with $\gamma/\mu = 1000$ are then compared in estimating $\langle Z_1Z_2\rangle$ for these circuits.
Exact state-vector evolution provides the reference value, and each method is evaluated using the capped relative error,
\begin{align}\label{eq:benchmark-plot}
    \min\{ 
      \frac{|\langle Z_1 Z_2\rangle_{\mathrm{approx}}-\langle Z_1 Z_2\rangle_{\mathrm{exact}}|}
           {|\langle Z_1 Z_2 \rangle_{\mathrm{exact}}|}
      , 1\}.
\end{align}

The results are shown in Fig.\;\ref{fig:benchmark}.
At the same bond dimensions, the proposed algorithm yields expectation value errors two to three orders of magnitude smaller than those of variational state compression.

\section{Conclusion}\label{sec:conclusion}
This work introduces an algorithm that sweeps along the time direction, jointly guiding state and observable approximations by how their errors combine in the final expectation value.
In the tested settings, the proposed algorithm yields substantially smaller errors than the conventional methods.
The loss functions considered here are heuristic choices whose optimal form remains open.
Further exploration of this direction may improve the accuracy and efficiency of dynamical expectation value estimation.


\newpage

\bibliographystyle{unsrt}
\bibliography{reference}


\newpage
\appendix
\section{Detailed derivation of the error decomposition}\label{sec:appendix-error-relation}
This appendix derives the exact error decomposition in Eq.~\eqref{eq:error}.
For a circuit of $L$ unitary gates, the exact state and observable at time slice $t$ are
\begin{align*}
    \rho_t
&=
    U_t\cdots U_1 \rho_0 U_1^\dagger\cdots U_t^\dagger,
    \\
    O_t
&=
    U_{t+1}^\dagger\cdots U_L^\dagger O U_L\cdots U_{t+1},
\end{align*}
for $t=0,\ldots,L$, with empty products interpreted as the identity.
Following the notation of Section~\ref{sec:rationale}, let
$\tilde{\rho}_t^{(i)}$ and $\tilde{O}_t^{(i)}$ denote the approximate state and observable,
where $i=0$ labels the initial SVD-based approximations and $i=1$ labels those obtained
after the forward and backward sweeps in Algorithm~\ref{alg:main-algorithm}.
For both values of $i$, the endpoints satisfy
$\tilde{\rho}_0^{(i)}=\rho_0$ and $\tilde{O}_L^{(i)}=O$.
The corresponding estimate of the expectation value is
\begin{equation*}
    b^{(i)} = \operatorname{Tr}\!\left[\rho_0 \tilde{O}_0^{(i)} \right].
\end{equation*}
The exact expectation value is
\begin{equation*}
    E = \operatorname{Tr}\!\left[\rho_t O_t \right],
\end{equation*}
which is independent of $t$ by unitary evolution and the cyclicity of the trace.
Thus, the error in $b^{(i)}$ is
\begin{equation*}
    E - b^{(i)}
    =
    \operatorname{Tr}\!\left[\rho_0 \left(O_0 - \tilde{O}_0^{(i)} \right) \right].
\end{equation*}
To express this error as a sum of local contributions, define
\begin{equation*}
    q_t^{(i)}
    :=
    \operatorname{Tr}\!\left[
        \tilde{\rho}_t^{(i)}
        \left(O_t - \tilde{O}_t^{(i)} \right)
    \right].
\end{equation*}
This auxiliary quantity evaluates the observable error in the approximate state.
The endpoint conditions give
\begin{equation*}
    q_0^{(i)} = E-b^{(i)},
    \qquad
    q_L^{(i)} = 0,
\end{equation*}
so summing successive differences gives the telescoping identity
\begin{align*}
    \sum_{t=1}^{L} \left( q_{t-1}^{(i)} - q_{t}^{(i)} \right)
&=
    q_0^{(i)} - q_L^{(i)}
    =
    E-b^{(i)}.
\end{align*}
The differences retain their signs; no monotonicity of $q_t^{(i)}$ is assumed.

Define the backward propagation defect at time slice $t$ by
\begin{equation*}
    d_t^{(i)}
    :=
    U_{t+1}^\dagger\tilde{O}_{t+1}^{(i)}U_{t+1}
    -
    \tilde{O}_t^{(i)},
    \qquad t=0,\ldots,L-1.
\end{equation*}
This defect is the difference between the exact one-step propagation of the stored
observable and its compressed approximation. It is distinct from the accumulated
observable error $O_t-\tilde{O}_t^{(i)}$.
Using $O_{t-1}=U_t^\dagger O_tU_t$, this accumulated error satisfies
\begin{equation*}
    O_{t-1} - \tilde{O}_{t-1}^{(i)}
    =
    d_{t-1}^{(i)}
    +
    U_t^\dagger \left(O_t - \tilde{O}_t^{(i)}\right) U_t,
    \qquad t=1,\ldots,L.
\end{equation*}
Similarly, define the forward propagation defect by
\begin{equation*}
    p_t^{(i)}
    :=
    U_t \tilde{\rho}_{t-1}^{(i)} U_t^\dagger
    - \tilde{\rho}_t^{(i)},
    \qquad t=1,\ldots,L.
\end{equation*}
Substituting the recursion for the observable error into the definition of
$q_{t-1}^{(i)}$ and using the cyclicity of the trace yields
\begin{align*}
    q_{t-1}^{(i)}
    &= \operatorname{Tr}\!\left[
        \tilde{\rho}_{t-1}^{(i)} d_{t-1}^{(i)}
    \right]
    +
    \operatorname{Tr}\!\left[
        \tilde{\rho}_{t-1}^{(i)} U_t^\dagger
        \left( O_t - \tilde{O}_t^{(i)}\right) U_t
    \right]
    \\
    &= \operatorname{Tr}\!\left[
        \tilde{\rho}_{t-1}^{(i)} d_{t-1}^{(i)}
    \right]
    +
    \operatorname{Tr}\!\left[
        U_t \tilde{\rho}_{t-1}^{(i)} U_t^\dagger
        \left( O_t - \tilde{O}_t^{(i)}\right)
    \right]
    \\
    &= \operatorname{Tr}\!\left[
        \tilde{\rho}_{t-1}^{(i)} d_{t-1}^{(i)}
    \right]
    + q_t^{(i)}
    +
    \operatorname{Tr}\!\left[
        p_t^{(i)}
        \left( O_t - \tilde{O}_t^{(i)}\right)
    \right].
\end{align*}
Rearranging gives the local identity
\begin{equation}
    \label{eq:appendix-q-difference}
    q_{t-1}^{(i)} - q_t^{(i)}
    =
    \operatorname{Tr}\!\left[
        \tilde{\rho}_{t-1}^{(i)} d_{t-1}^{(i)}
    \right]
    +
    \operatorname{Tr}\!\left[
        p_t^{(i)}
        \left( O_t - \tilde{O}_t^{(i)}\right)
    \right].
\end{equation}
Summing Eq.~\eqref{eq:appendix-q-difference} over $t=1,\ldots,L$ and using
the telescoping identity above gives
\begin{align}
    \label{eq:appendix-decomposition}
    E - b^{(i)}
&=
    \sum_{t=0}^{L-1}
    \operatorname{Tr}\!\left[
        \tilde{\rho}_t^{(i)} d_t^{(i)}
    \right]
    +
    \sum_{t=1}^{L-1}
    \operatorname{Tr}\!\left[
        p_t^{(i)}
        \left( O_t - \tilde{O}_t^{(i)}\right)
    \right].
\end{align}
The second sum ends at $L-1$ because $O_L-\tilde{O}_L^{(i)}=0$.
The first sum includes all $L$ backward defects, including $d_{L-1}^{(i)}$,
which need not vanish. Empty sums are understood to be zero.
The two sums in Eq.~\eqref{eq:appendix-decomposition} are $R_1^{(i)}$ and
$R_2^{(i)}$, respectively.

To recover the expression for $R_2^{(i)}$ used in the main text, define the
accumulated state and observable errors by
\begin{equation*}
    H_t^{(i)}:=\rho_t-\tilde{\rho}_t^{(i)},
    \qquad
    D_t^{(i)}:=O_t-\tilde{O}_t^{(i)},
    \qquad t=0,\ldots,L.
\end{equation*}
Subtracting the approximate state from the exact propagation relation and using
the definition of $p_t^{(i)}$ gives
\begin{align*}
    H_t^{(i)}
    &= U_t\rho_{t-1}U_t^\dagger-\tilde{\rho}_t^{(i)}
    \\
    &= U_t\left(\rho_{t-1}-\tilde{\rho}_{t-1}^{(i)}\right)U_t^\dagger
       +p_t^{(i)}
     = U_t H_{t-1}^{(i)}U_t^\dagger+p_t^{(i)}.
\end{align*}
The observable-error recursion derived above becomes
\begin{equation*}
    D_{t-1}^{(i)}=d_{t-1}^{(i)}+U_t^\dagger D_t^{(i)}U_t.
\end{equation*}
Both recursions hold for $t=1,\ldots,L$, with $H_0^{(i)}=0$ and $D_L^{(i)}=0$.
Thus, each accumulated error is a sum of propagated local defects.
To write these sums explicitly, let
\begin{equation*}
    U_{t:s}:=U_t\cdots U_{s+1},
    \qquad U_{s:s}:=I,
    \qquad 0\le s\le t\le L,
\end{equation*}
denote the unitary propagation from time slice $s$ to time slice $t$.
Iterating the two recursions yields
\begin{align*}
    H_t^{(i)}
    &=\sum_{s=1}^{t}U_{t:s}p_s^{(i)}U_{t:s}^\dagger,
    \\
    D_t^{(i)}
    &=\sum_{s=t}^{L-1}U_{s:t}^\dagger d_s^{(i)}U_{s:t}.
\end{align*}
Substituting the second expression into the remainder in
Eq.~\eqref{eq:appendix-decomposition}, using the cyclicity of the trace, and
exchanging the order of the finite sums gives
\begin{align*}
    R_2^{(i)}
    &=\sum_{t=1}^{L-1}\operatorname{Tr}\!\left[p_t^{(i)}D_t^{(i)}\right]
    \\
    &=\sum_{t=1}^{L-1}\sum_{s=t}^{L-1}
      \operatorname{Tr}\!\left[
          U_{s:t}p_t^{(i)}U_{s:t}^\dagger d_s^{(i)}
      \right]
    \\
    &=\sum_{s=1}^{L-1}\operatorname{Tr}\!\left[
          \left(\sum_{t=1}^{s}U_{s:t}p_t^{(i)}U_{s:t}^\dagger\right)d_s^{(i)}
      \right]
    \\
    &=\sum_{s=1}^{L-1}\operatorname{Tr}\!\left[H_s^{(i)}d_s^{(i)}\right].
\end{align*}
In the last step, the inner sum is precisely $H_s^{(i)}$.
Since $H_0^{(i)}=0$, the lower limit can be extended to zero.
Substituting the definition of $H_s^{(i)}$ and relabeling the summation index gives
\begin{equation}
    \label{eq:appendix-R2}
    R_2^{(i)}
    =
    \sum_{t=0}^{L-1}
    \operatorname{Tr}\!\left[
        \left(\rho_t - \tilde{\rho}_t^{(i)}\right) d_t^{(i)}
    \right].
\end{equation}
The two expressions for $R_2^{(i)}$ are therefore equal as sums; their individual
terms need not coincide. Together, Eqs.~\eqref{eq:appendix-decomposition} and
\eqref{eq:appendix-R2} establish the decomposition
$E-b^{(i)}=R_1^{(i)}+R_2^{(i)}$ stated in the main text.
The derivation uses only the propagation relations and endpoint conditions;
it does not require the local compression problems to be solved exactly.

\end{document}